# Distributed Elections In An Archimedean Ring Of Processors*

(Preliminary Version)

Paul M.B. Vitányi

*Centre for Mathematics and Computer Science (C.W.I.), Amsterdam†*

Summary

[16th ACM Symposium on Theory of Computing, Washington D.C., 1984, 542 - 547]

Unlimited asynchronism is intolerable in real physically distributed computer systems. Such systems, synchronous or not, use clocks and timeouts. Therefore the magnitudes of elapsed absolute time in the system need to satisfy the axiom of Archimedes. Under this restriction of asynchronicity logically time-independent solutions can be derived which are nonetheless better (in number of message passes) than is possible otherwise. The use of clocks by the individual processors, in elections in a ring of asynchronous processors without central control, allows a deterministic solution which requires but a linear number of message passes. To obtain the result it has to be assumed that the clocks measure finitely proportional absolute time-spans for their time units, that is, the magnitudes of elapsed time in the ring network satisfy the axiom of Archimedes. As a result, some basic subtleties associated with distributed computations are highlighted. For instance, the known nonlinear lower bound on the required number of message passes is cracked. For the synchronous case, in which the necessary assumptions hold *a fortiori,* the method is -asymptotically- the most efficient one yet, and of optimal order of magnitude. The deterministic algorithm is of -asymptotically- optimal bit complexity, and, in the synchronous case, also yields an optimal method to determine the ring size. All of these results improve the known ones.

> ... since the centre of the sphere has no magnitude, we cannot conceive it to bear any ratio whatever to the surface of the sphere.
>
> Archimedes, *The Sand-Reckoner*

## 1. Introduction

We address the issue of time in distributed systems. Under genuinely reasonable assumptions about time in distributed systems there exist, for some problems, logically time-independent solutions which are more efficient than achievable with unlimited asynchronism. The point here is that an algorithm can be robust enough to function under any assumption whatever about time in the system, but its efficiency may in a nontrivial sense change with the assumption. The solution for the distributed election problem in ring networks described below should be taken to illustrate this thesis rather than as a serious proposal for crash recovery in token ring networks. Its message pass complexity will be shown O($sN$), with $s$ a measure of the asynchronicity in the system and $N$ the number of processors, in contrast to the optimal $\Theta(N \log N)$ solution for the unlimited asynchronous case. So consider a set of processors, arranged in a circle. Each processor has a unique name, say a positive integer. Apart from this, the situation for the processors is symmetrical. Communication between processors occurs only between neighbors around the circle. There are $N$ processors, but this is not known to the processors themselves. It is a common logical organization of a network of processors to locate them on such a (physical or virtual) ring. A natural feature of crash recovery in computer networks, or other network tasks where there is no central control, consists in first reaching unanimous agreement on the choice of a unique leader. For example, in a token ring network, where the token is lost or multiplied, a *single* new token has to be created. Thus, following some initial, possibly local, disturbance observed by at least one process, the distributed processes need to find an extremum on which they all agree. The problem is treated in [Le, CR, HS, Fr, Bu, Ga, DKR, PKR, IR]; ring networks in general in e.g. [DSM, St, SPC, Ta]. Elections appear to be a key problem since the number of message passes one has to expend, in order to reach any agreement whatever in a decentralized network, seems to be at least that required by leader finding, and usually not of greater order of magnitude (because after a leader is agreed upon the remainder is not too costly).

*Previous Solutions for Elections in Asynchronous Rings.* In an asynchronous ring there is no global clock for synchronizing the actions. Moreover, arbitrarily long delays may occur between the sending and receiving of a message. Still, all such delays are finite. The easiest election strategy is to have each processor, which becomes aware that an election is on, send a signed message around the circle in one direction. If messages of lower indexed processors are not passed on by higher indexed processors then the only message returning to its origin is that of the highest indexed processor [Le]. This takes $\Theta(N^2)$ message passes in the worst case. In [Fr] a method with bidirectional message passing is given using a worst case amount of shown that the problem requires $\Omega(N \log N)$ message passes. Since the methods of [Fr], also [HS, Bu], use $O(N \log N)$ message passes, they are therefore considered to be asymptotically optimal to within a constant multiplicative factor. The Le Lann method [Le] is superior in the sense that it operates by passing messages in *one direction* only. However, in [DKR] a one directional solution is proposed with $O(N \log N)$ message passes in the worst case. Since in [PKR] the $\Omega(N \log N)$ lower bound is obtained on the *average* number of message passes needed to solve the problem in the asynchronous one directional case, the matter seemed wholly resolved.

---

\* This work was supported by the *Stichting Mathematisch Centrum.*

† Full Address: Centre for Mathematics and Computer Science (C.W.I.), Kruislaan 413, 1098 SJ Amsterdam, The Netherlands.

*Previous Solutions for Elections in a Synchronized Ring.* In a synchronized ring there is a global clock, or some other device, which coordinates the actions in the individual processors so that they proceed in lock-step. The communication delay between the sending and receiving of a message is a priori bounded in terms of time units of the global clock. Probabilistic algorithms have been proposed [IR] for solving the election problem in linear time on the average, provided the size of the ring is known and the processes are synchronous (with communication delay zero). There is no nontrivial lower bound for the average number of messages in the synchronous version when the size of the ring is not known, nor for the general case where the size of the ring is known.

*Improved Solutions using Time and Clocks.* The purpose here is to find a better way, by using clocks, for solving the decentralized election problem for asynchronous ring networks, which cracks the established lower bound in [Bu, PKR]. Despite the simplicity of the method, all results below improve the known ones.

*Asynchronous Case.* To achieve the deterministic one-directional solution with a linear number of message passes, the concept of asynchronicity has to be restricted to what may be called Archimedean asynchronicity. Unrestricted asynchronicity, it will be argued, is too harsh an environment for the questions at issue. That is, the $\Omega(N \log N)$ lower bound is established in [Bu, PKR] under assumptions so hostile that they preclude a usable solution anyway. In addition, the proposed solution has an optimal bit complexity. It may need message queues.

Solutions for distributed control problems usually do not use clocks and time and make no assumptions about relative time rates. This, in order to rule out constructions that depend on timing for their correct operation. The *message pass* complexity measure to determine the better one of two solutions is a consequence of this expulsion of time. Sometimes time is introduced afterwards to determine the running time of a logically time-independent procedure. The *correctness* and *termination* of the solution below is independent of the timing assumptions. The *message pass* complexity and the *bit* complexity depend on the use of time and clocks and are better the more synchronous the system behaves. The presented solution uses $O(sN)$ message passes and a correspondingly efficient number of bits. The coefficient $s$ is a scaling factor which measures the asynchronicity of the system. It can be eliminated by the use of appropriate parameters in the Protocols. Contrast this with the known $\Omega(N \log N)$ lower bound on the average number of message passes for the non-Archimedean case. In Section 4 we shall express the running time complexity of the solution in the *walk time* of the ring, that is, the time for a single bit to circle the

*Synchronous Case.* The deterministic solution presented below is outright superior, viz. runs in a linear number of message passes, for synchronous systems, for such systems are *a fortiori* Archimedean ($s=1$). The bit complexity is also optimal. The method can be used to determine the unknown ring size in optimal complexity in message passes and passed bits. (Optimal in the sense of order of magnitude.) In the synchronous case the method does not need unbounded message queues.

## 2. Distributed systems and Archimedean time

In asynchronous distributed systems it is usually assumed that each processor has its own clock. Although it may have been explicitly stated that these clocks are not synchronized, it is usually either implied or stated in plain words that, although these clocks do not indicate the same time, there is some proportion between elapsed time spans. That is, if an interval of time has passed on the clock for processor *A*, a proportional period of time has passed on the clock for processor *B*. This assumption allows us to challenge the $\Omega(N \log N)$ lower bound on the required number of message passes in [Bu, PKR].

We can express the assumption by stating that in the type of asynchronous network we consider, the magnitudes of elapsed time satisfy the axiom of Archimedes. The axiom of Archimedes holds for a set of magnitudes if, for any pair $a$, $b$ of such magnitudes, there is a multiple $na$ which exceeds $b$ for some natural number $n$. It is called Archimedes' axiom* possibly due to application on a grand scale in *The Sand-Reckoner*.

We assume that the magnitudes of elapsed time, for instance as measured by local clocks amongst different processors or by the clock of the same processor at different times, as well as the magnitudes consisting of communication delays between the sending and receiving of messages, measured in for instance absolute physical time, all together considered as a set of magnitudes of the same kind, satisfy the Archimedean axiom. In physical reality it is always possible to replace a magnitude of elapsed time, of any clock or communication delay, by a corresponding magnitude of elapsed absolute physical time, thus obtaining magnitudes of the same kind. Purists may throw in relativistic corrections. We assume a global absolute time to calibrate the individual clocks; using relative time by having the clocks send messages to one another yields the same effect - for the purposes at hand. If we do not restrict ourselves, so to speak, to Archimedean distributed systems, then the processors in the system may not have any sense of time or have clocks which keep purely subjective time, so that the unit time span of each processor is unrelated to that of another. That is, the set of time

---

\* In *Sphere and Cylinder* and *Quadrature of the Parabola* Archimedes formulates the postulate as follows. ''The larger of two lines, areas or solids exceeds the smaller in such a way that the difference, added to itself, can exceed any given individual of the type to which the two mutually compared magnitudes belong''. The axiom appears earlier as Definition 4 in Book 5 of Euclid's *Elements* which elaborates the work on proportion of Eudoxus of Knidos (408 BC - 355 BC): ''Magnitudes are said to have a ratio to one another, which are capable, when multiplied, of exceeding one another''. The Archimedean axiom, together with Definition 5 in [Op. cit.], yields the complete theory of proportion for kinds of magnitudes that have a ratio to one another. It also figures prominently in the limit arguments of Eudoxus' exhaustion method.

units is non-Archimedean by the length of every time unit not being less than a finite times that of any other in the absolute global time scale; or the communication delays having no finite ratio among themselves or with respect to subjective processor clocks. As a consequence Extrema Finding or any other type of synchronization in a deterministic fashion becomes impossible. For consider:

 -Any process, pausing indefinitely long with respect to the time-scale of the others, between events like the receiving and passing of a message, and also any infinite communication delay, effectively aborts an election in progress. A process can never be sure that it is the only one which considers itself elected.

 -Without physical time and clocks there is no way to distinguish a failed process from one just pausing between events.

 -A user or a process can tell that a system has crashed only because he has been waiting too long for a response.

The nature of time and clocks in distributed systems is discussed in detail in [Le, La, Ga], where the notion of a distributed system, in which elections as described are at all possible, agrees with that of an Archimedean distributed system as defined below. Distributed systems in the sense of *physically* distributed computer networks communicate by sending signed messages and setting timers. If an acknowledgement of safe receipt by the proper addressee is not received by the sender before the timer goes off, the sender sends out a new copy of the message and sets a corresponding timer. This process is repeated until either a proper acknowledgement is received or the sender concludes that the message cannot be communicated due to failures. Thus, clocks and timeouts seem necessary attributes of real distributed systems [Ta] and non-Archimedean time in the system is intolerable outright.

*Definition*. A distributed system is *Archimedean* from time $t_1$ to time $t_2$ if the ratio of the time intervals between the ticks of the clocks of any pair of processors, and the ratio between the communication delay between any adjacent pair of processors and the time interval between the ticks of the clock of any processor, is bounded by a fixed integer during the time interval from $t_1$ to $t_2$.

## 3. Decentralized leader finding using clocks

*Asynchronous Case.* The basic feature of all efficient solutions for the decentralized election problem is how to eliminate future losers and the messages they send fast enough. The matter is complicated by the symmetry of the individual processors in the ring; hence the $\Omega(N \log N)$ lower bound on the number of message passes. Yet the situation for the individual processors is not entirely symmetrical, since they have unique names. (For a ring consisting of wholly identical processors deterministic leader finding is impossible, since the situation is symmetrical for each processor.) In previous solutions the unique names are used in the selection process to shut off losing processors or to eliminate their messages. Rather than using names only in comparisons, we can also use them to restrict the number of message passes of messages originated by future losers. To achieve this, we use time and clocks. Assume that each processor has its own clock and that the absolute time span that elapses between the ticks of any clock, together with the greatest communication delay any other clock. By setting that fixed multiple to $\lceil u/m \rceil$, where $u/m$ is the ratio between the first mentioned time interval $u$ and the second one $m$, for the given clocks and communication delays, we see that the assumption holds for Archimedean rings of processors.

The algorithm is basically a souped-up version of Le Lann's method. Initially all processors are functioning happily in their normal mode which we, for the present purposes, call being *asleep*. Suddenly, one or more *awake*, that is, become aware that an election is due. Between this time and the time the Elected One is determined, and all processors have been notified thereof, any processor which awakes executes the Protocol below. Processes awake spontaneously, and in any event when they receive a wakeup message from their anticlockwise neighbor. On notification of a successful election by a *sleepwell* message a process falls asleep again. We give the Protocol, explain the method, prove it correct and analyse its complexity.

**Protocol to be executed when processor *i* awakes.**

Send *wakeup* message to clockwise neighbor; Set $k$ equal to $i$ and set *timer* equal to 1;

REPEAT IN EACH (LOCAL) TIME ''UNIT'':
Read incoming message $M$ from anticlockwise neighbor (if no message is received in this time unit then assume $M = M_j$ with $j > i$);

**if** ''I am asleep'' and $M$ is the *sleepwell* message **then** the election is finished; #Everyone knows the winner is me, that is, *i*. The *sleepwell* message need not contain the name of the Elected One.#

**if** ''I am awake'' and $M$ is the *sleepwell* message **then**
**begin**
Elected One $\leftarrow k$;
   send *sleepwell* message to clockwise neighbor and go asleep
**end**

**if** ''I am awake'' and $M = M_j$ is an *election* message **then**
**begin**
**if** $j = k$ **then**
   **begin**
   Elected One $\leftarrow k$; #$k = i$#
      send *sleepwell* message to clockwise neighbor and go asleep
   **end**

**if** $j < k$ **then begin** $k \leftarrow j$; *timer* $\leftarrow f(k)$ **end**
**if** $j > k$ **then**
**begin**
*timer* $\leftarrow$ *timer* $-1$;
   **if** *timer* $= 0$ **then** send $M_k$, containing $k$, to clockwise neighbor
**end**
**end**

**Figure.** Election Protocol.

Subsequent to the initial prodding of any processor, in $N$ message passes around the ring, all processors are aware that an election is in progress. This is encouched in the Protocol as follows. Each processor can be *asleep* or *awake*. If a processor changes its state from *asleep* to *awake* it sends a *wakeup* message to its clockwise neighbor; a processor changes its state from asleep to awake either because it receives a wakeup mes-awake it knows that an election is in progress. In precisely $N$ message passes of *wakeup* messages all processors in the ring are awake. The *wakeup* message can consist of a single bit. Now recall that all processors are supposed to have a unique name, which can be interpreted as a positive integer. Following the *wakeup* message emission, each processor $i$ generates a single *election* message $M_i$. The Protocol states that a message $M_i$, originating from processor $i$, waits $f(i)$ of the local time units, of the processor which received it, before being transmitted to the clockwise next processor. Assume that $f$ is a monotone strictly increasing function. Each *election* message $M_i$ containing $i$ is preceded by a wakeup signal also originating from processor $i$. Thus, with respect to the election campaign, all processors are effectively awake, as soon as one of them is awake. During the campaign, whenever a message with a higher number meets a lower numbered processor, that message is annihilated. Whenever a lower numbered message overtakes a higher numbered message, it annihilates the latter. Hence, all messages -but its own- are annihilated by the lowest numbered processor and the lowest numbered message annihilates all other messages when it overtakes them. So all messages have been smashed between hammer and anvil by the time the lowest numbered message returns to its origin, leaving it the only one in the ring. It immediately follows that the algorithm is correct. It remains to estimate its complexity. Globally and absolutely speaking, $u$ is an upper bound on the lengths of the individual time units increased with the largest communication delay, and $m>0$ is a lower bound on the length of the individual time units. Let, furthermore, the least name of a processor be $l$. Then the message $M_l$ needs no more than $Nf(l)u$ absolute time to make the tour around the ring of processors. Subsequently, $l$ sends a special *sleepwell* message around, informing the other processors it is the elected one. The *sleepwell* message circles the ring at top speed, so it takes no more than $Nu$ absolute time. This message need not contain index $l$, since message $M_l$ has passed all processors in the ring and therefore set all local variables $k$ to $l$. Thus, the *sleepwell* message can consist of but a few bits. Following the original prodding, in $N$ message passes and in no more than $Nu$ absolute time, all processors are awake. In the course of these events, an election message $M_i$ can, during its allotted time, engage in no more than

$$\frac{Nu(f(l)+1)}{mf(i)} \qquad (1)$$

message passes. Hence, the total number of message passes in the system is not greater than:

$$2N + \frac{Nu(f(l)+1)}{m} \sum_{i \in I} \frac{1}{f(i)} \ , \qquad (2)$$

where $I$ denotes the set of processor names. Thus, for $f(i) \geq 2^i$, the sum converges to something between $1/f(l)$ and $2/f(l)$. Consequently, the number of message passes in the system is bounded above by $2N+3Nu/m$ ($l \geq 1$). Assuming that $u/m$ does not depend on $N$, the method yields a linear upper bound on the number of message passes in the system. Alternatively, we can eliminate $u/m$ from the upper bound by incorporating it in $f$, for instance, by choice of $f=(u2/m)^i$ or larger. See Section 4 for more discussion on this topic.

Separating the effects of the clock delays and the interprocessor signal propagation delays yields the following. Let $u'$ stand for clocks. Let the combined interprocessor signal propagation delay around the ring be $w_s$. Then $Nu \geq Nu'+w_s$. If there is some quality control in the clock factory, so that $u'-m < \varepsilon$ for some fixed $\varepsilon$, then a statistically sound assumption is to distribute the clock delays *homogeneously* over the interval $[u', m]$, and $u'/m < 1+\varepsilon/m$. This approach yields equations analogous to (1) and (2) and a similar result. In (1) we add $2w_s$ to the numerator and $w_s$ to the denominator, and replace $u$ by $u'$. The resulting message pass complexity turns out to be less than $7N+3\varepsilon N/m$.

Another measure of interest is the total number of *bits* passed in the system. In previous solutions the way of encoding the signature $i$ in a message $M_i$ did not matter very much. Any scheme using $\log N$ bits sufficed. In the present solution though, we can take advantage of the fact that large messages are not passed often. Thus, we code the signature $i$ of $M_i$ in *dyadic* numbers without leading zeroes. Recall, that dyadic numbers use the digits 1 and 2, with the normal binary weight in their respective positions, instead of the customary digits 0 and 1, and 1, 2, 3, 4, 5, 6, $\cdots$ are encoded as 1, 2, 11, 12, 21, 22, $\cdots$. By the argumentation above, and assuming that the message $M_i$ contains but $O(\log i)$ bits, by dyadic encoding, the total number of bits passed in the system in the sketched strategy is bounded by

$$2N + \frac{Nu(f(l)+1)}{m} \sum_{i \in I} \frac{\log i}{f(i)} \ . \qquad (3)$$

Similar to above, for $f(i) \geq 2^i$, the sum converges to $c' \log l/f(l)$ for some constant $c'$, and the total number of bits passed is bounded above by $cNu \log l/m$ for some small constant $c$.

## 4. A closer look

*The worst case performance.* The $O(Nu/m)$ upper bound on the number of message passes of the solution is *linear* in $N$ and $u/m$. If compelled by practical considerations and accompanying quality control to consider only networks with $u/m < s$, where $s$ is some fixed constant, then the number of message passes is truly $O(N)$. The assumption of Archimedean time and clocks in the system has enabled us to use the *names* of the processors in a new way to cut down on the number of message passes. The implied slack with the known $\Omega(N \log N)$ lower bound on the number of message passes for the unlimited asynchronous case is taken up by the asynchronicity factor $u/m$ which is by its nature independent of $N$. It seems contrived to suppose that $u/m$ rises unboundedly with $N$. Even if we do suppose this to be the case then the factor $u/m$ can be eliminated by incorporation in $f$ as follows. (Incorporating $u/m$ in $f$ has the drawback of implicitly using a global system parameter in the Protocol.) The winning message $M_l$ makes precisely $N$ message passes. Therefore, we can replace the upper bound (2) on the number of message passes by

$$3N + \frac{Nu(f(l)+1)}{m} \sum_{i \in I-\{l\}} \frac{1}{f(i)} \ , \qquad (4)$$

which for, e.g., $f(i)=(2u/m)^i$ yields no more than $3N+N(1+1/f(l)) < 5N$ ($l \geq 1$) message passes. Similarly, the number of passed bits is, for this choice of $f$, bounded above by $2N+3N \log l$. Thus, the number of passed bits is linear in $N$, if in $f$) also $l$ is independent of $N$. If $l$ would depend on $N$ at all then it seems more natural to suppose that it *decreases* with $N$. (If we add a new processor to the system then we choose a new name for the new processor only and not for all the old ones.) The problem requires $\Omega(N \log l)$ passed bits in any case, since the name of processor $l$ has to be communicated to all processors. The *time complexity* of the above procedure is, for $f(i) = 2^i$, no more than $Nu(2^l+2)$, which is pretty good if $l$ is reasonably low, like 1.

Note that *any* $f$ such that $\lim_{i \to \infty} i^\varepsilon / f(i) = 0$, for some $\varepsilon > 1$, gives asymptotically similar results. The message pass complexity for such $f$ is $O(Nul/m)$ since $\sum_{i \geq l} l^\varepsilon / i^\varepsilon \in O(l)$.

*Synchronous case.* In the synchronous case the above *deterministic* solution yields the various stated asynchronous upper bounds with $u = m$. This without any assumptions whatever, since synchronous systems are *a fortiori* Archimedean. Since all of the resulting bounds are *linear* in $N$ and within a small multiplicative constant of the trivial lower bounds, for the respective measures, the solution is optimal. By counting time, as part of the Protocol of each processor, the network can determine the unknown ring size $N$ in the extreme processor $l$ using a total of $O(N)$ message passes and $O(N \log l)$ passed bits.

*The Worst-Case Performance under adversary scheduling with fixed f.* If we assume that $f$ is fixed and the system can be adjusted then the worst what can happen by adversary scheduling both the unit delays of all processors and the processor placement around the ring is square in $N$. Let the unit delay of processor $i$ be $u_i = 2^{N-i+1}$ and $f(i) = 2^i$. Place furthermore the processors, in ascending order, clockwise around the ring. Thus, 1 is the clockwise neighbor of $N$ and $i+1$ the clockwise neighbor of $i$, $1 \leq i < N$. Under these conditions, no message can overtake another one, so all messages are annihilated by processor 1. So message $M_i$ makes $N-i+1$ message passes leading to $N(N+2)/2$ message passes altogether. This is essentially the case covered in [Le, CR]. This shows that the upper bound estimate in the last section is too crude, since it exceeds this bound by choice of $u/m \in \Omega(N^{1+\varepsilon})$ for all $\varepsilon > 0$.

*The Average-Case Performance.* In [CR] the *expected* number of message passes over all possible permutations of the processors over the ring is considered. They find $O(N \log N)$. We will do the same for the method described under the assumption that each permutation of names of processors over the ring has the same probability. We do not need to assume anything about the distribution of the delays. The *walk time* $w = w_p + w_s$ consists of the combined 1 bit per station delay $w_p$ plus the signal propagation delay $w_s$ over the entire ring [DSM, St, Stu, Ta]. More precisely, the walk time of a token ring network is the time it takes for a single bit to circumnavigate an empty ring. It has two components: the propagation time on the cable, about 5 nanoseconds per meter of cable, and the node delays. Each node has 1 or more bits of storage. In effect, the node buffers are like a big distributed shift register. At every clock tick, all bits shift one position. Each node needs at least one bit of delay so it can inspect the last bit of the token and change it to remove the token, if need be. A current ring may have 5 bits per node, so a short two node network will have enough bits to hold a token with a little room to spare. In a short network with 1 bit of delay per node, a two node network would be too small to store a token. The walk time is independent of the message length, and function of the cable length, number of delay bits per node, and the transmission speed (the reciprocal of which is how often the big shift register is advanced). Thus, a one-bit message circles the entire ring in $w$ absolute time. An $i$-bit *election message* takes in the order of $w_s + w_p f(i) \log i$ absolute time, since in our solution we assume that all bits of the messages are read by the processors in the ring and acted upon before release. The expected number of message passes of *election message* $M_i$ is found by dividing the maximal available time $O(w + w_s + w_p f(l) \log l)$ by the time $\Omega(w_s + w_p f(i) \log i)$ for $M_i$ to circumnavigate the entire ring and multiplying this fraction with the total number $N$ of passes around the ring. Reasoning analogous to before, the expected number of message passes in the ring is therefore not greater than of order

$$2N + N \sum_{i \in I} \frac{w + w_s + w_p f(l) \log l}{w_s + w_p f(i) \log i} \ . \tag{5}$$

This is, for $f(i) \geq 2^i$ and $l \geq 1$, of $O(Nw/w_p)$, or more precisely of $O(N(1 + (w/(w_p f(l) \log l)))$. If we assume that the communication delays are negligible, or $w/w_p$ is a constant independent of $N$, or $f(i) \geq w 2^i / w_p$, then the expected number of message passes is $O(N)$.

*Minimal Time Performance.* If, instead of the number of message passes in the system, we want to minimize the *absolute time* for the solution, then the previous message-pass optimal solutions in the references will all do pretty poorly when we consider adversary scheduling of delays, processor names and wake-up moments around the ring. The solution given above will take time not greater than $2w + w_s + w_p f(l) \log l$. By a simple variant we can eliminate the factor $f(l)$. Choose $f$, depending on both the processor $P_i$ and the entrant message $M_j$, as $f(i, j) = \lfloor 2^{j-i} \rfloor$ in the Protocol. Then the winning election message $M_l$ takes precisely $w_s + w_p \log l$ absolute time to circle the ring. Therefore, the solution time is not greater than $3w + w_p(\log l - 1)$. This is a reversion to the method in [CR] and reaches virtually the trivial lower bound on the absolute running time, but uses $\Theta(N^2)$ message passes in the worst case, and $\Theta(N \log N)$ on the average. By choice of $f$ in the above Election Protocol we can optimize different complexity measures separately; can we also optimize them simultaneously?